\documentclass{bmcart}

\usepackage[utf8]{inputenc} 

\usepackage{graphicx}
\usepackage{amsmath,environ}

\startlocaldefs
\endlocaldefs

\begin{document}

\begin{frontmatter}

\begin{fmbox}
\dochead{Research}


\title{Assessing the Interplay between travel patterns and SARS-CoV-2 outbreak in realistic urban setting}


\author[
   addressref={aff1},
   email={rohan.patil@iitgn.ac.in},
   noteref={n1}
]{\inits{RP}\fnm{Rohan} \snm{Patil}}
\author[
   addressref={aff2},
   email={dave_raviraj@iitgn.ac.in},
   noteref={n1}
]{\inits{RD}\fnm{Raviraj} \snm{Dave}}
\author[
   addressref={aff1},
   email={harsh.patel@iitgn.ac.in}
]{\inits{HP}\fnm{Harsh} \snm{ Patel}}
\author[
   addressref={aff3},
   email={viraj.shah@iitgn.ac.in}
]{\inits{VS}\fnm{Viraj M.} \snm{ Shah}}
\author[
   addressref={aff4},
   email={deepchakrabarti.19@gmail.com}
]{\inits{DC}\fnm{Deep} \snm{Chakrabarti}}
\author[
   addressref={aff2},                   
   corref={aff1},                       
   email={bhatia.u@iitgn.ac.in}   
]{\inits{UB}\fnm{Udit} \snm{Bhatia}}


\address[id=aff1]{
  \orgname{Discipline of Computer Science and Engineering, Indian Institute of Technology}, 
  \city{Gandhinagar},                              
  \cny{India}                                    
}
\address[id=aff2]{
  \orgname{Discipline of Civil Engineering, Indian Institute of Technology}, 
  \city{Gandhinagar},                              
  \cny{India}                                    
}
\address[id=aff3]{
  \orgname{Discipline of Mechanical Engineering, Indian Institute of Technology}, 
  \city{Gandhinagar},                              
  \cny{India}                                    
}
\address[id=aff4]{
  \orgname{King George Medical University}, 
  \city{Lucknow},                              
  \cny{India}                                    
}

\address[id=aff5]{
  \orgname{Discipline of Physics, Indian Institute of Technology}, 
  \city{Gandhinagar},                              
  \cny{India}                                    
}


\begin{artnotes}
\note[id=n1]{Equal contributors} 
\end{artnotes}

\end{fmbox}


\begin{abstractbox}

\begin{abstract} 
\parttitle{Background}
The dense social contact networks and high mobility in congested urban areas facilitate the rapid transmission of infectious diseases. Typical mechanistic epidemiological models are either based on uniform mixing with ad-hoc contact processes or need real-time or archived population mobility data to simulate the social networks. However, the rapid and global transmission of the novel coronavirus (SARS-CoV-2) has led to unprecedented lockdowns at global and regional scales, leaving the archived datasets to limited use. 
\parttitle{Findings}
While it is often hypothesized that population density is a significant driver in disease propagation, the disparate disease trajectories and infection rates exhibited by the different cities with comparable densities require a high-resolution description of the disease and its drivers. In this study, we explore the impact of creation of containment zones  on travel patterns within the city.  Further, we use a dynamical network-based infectious disease model to understand the key drivers of disease spread at sub-kilometer scales demonstrated in the city of Ahmedabad, India, which has been classified as a SARS-CoV-2 hotspot. We find that in addition to the contact network and population density, road connectivity patterns and ease of transit are strongly correlated with the rate of transmission of the disease. Given the limited access to real-time traffic data during lockdowns, we generate road connectivity networks using open-source imageries and travel patterns from open-source surveys and government reports. Within the proposed framework, we then analyze the relative merits of social distancing, enforced lockdowns, and enhanced testing and quarantining mitigating the disease spread.\parttitle{Scope} Our results suggest that the declaration of micro-containment zones within the city with high road network density combined with enhanced testing can help in containing the outbreaks until clinical interventions become available. 
\end{abstract}




\end{abstractbox}
%

\end{frontmatter}



\section*{Introduction}
Modern history is witness to several infectious disease pandemics which have shaped our knowledge of their epidemiology, transmission, and management \cite{shearer2020infectious}. In the past 200 years, at least four strains of influenza, seven waves of cholera, tuberculosis, and the human immunodeficiency virus (HIV) have accounted for the deaths of nearly 100 million people \cite{cobey2020modeling,egedeso2020preventing}. Mathematical modelling of infectious diseases allows us to predict an epidemic accurately, recognize uncertainties, and quantify situations to identify possible worst-case situations that can guide public health planning and decision making \cite{longini1988mathematical,grassly2008mathematical}.  

Classically, mathematical models of infectious diseases were dependent on the classification of individuals on their epidemiological status based on their potential ability to host and transmit a pathogen: Susceptible, Infectious, and Recovered [SIR] \cite{keeling2001dynamics,keeling2009mathematical,anderson1991infectious}. The SIR model is the most fundamental epidemiological model that relies on calculating the proportional burden of each of these three classes, and the transitions between them. In the context of an epidemic, assuming no births or deaths in a population, the only two possible transitions are infection (movement from susceptible to infected) and recovery (movement from infected to remission). For the sake of simplicity, it can be assumed that the susceptibility is proportional to the prevalence of infection or the disease burden in the community, and recoveries occur at a constant rate \cite{begon2002clarification}. Estimating transmission and recovery, the epidemic progresses exponentially until the growth rate slows and the epidemic curve plateaus; following which eventually over time the epidemic cannot be sustained and is eradicated \cite{keeling2001dynamics}.
A failing of the traditional SIR model is its inability to account for spatial aspects of disease spread. The 2001 dynamics of foot and mouth disease in the United Kingdom were demonstrated by explicit individualised modelling as transmission between farms that would not have been possible by traditional SIR models alone. Such models pointed at localised depletion of susceptible contacts as the mechanism for the slowing of the epidemic \cite{keeling2001dynamics,ferguson2001transmission}. The HIV pandemic is characterised by a chronic infective state. The transmission of such a sexually transmitted infection is dependent on the host immune status, the infected individual’s viral load, and multiple social aspects like sexual practices of interactions between multiple structured risk groups within the population.  In such a situation, the variability in infectious state predicts the progression of the epidemic and stochasticity merits more complex modelling \cite{wearing2005appropriate}. Similarly, pandemics of influenza or flu-like illnesses can be accounted for by age and risk-structured models, which are explained by distinct risk groups for infection and fatality (age, health care workers, comorbid conditions), and the inherent nature of people to preferentially socialise with others of a similar age—a principle known as assortativity \cite{keeling2009mathematical,lloyd2003curtailing}.
More complex modelling focuses on the individual as a unit of the population describing individual interactions of each person in a population, in contrast to estimating the proportions of people with a certain disease status. The shift from a population-based model to an individual-based model is a powerful tool that helps to account for complex biologically and socially relevant interactions \cite{riley2007large,ferguson2006strategies}.

Reliability of projections from individual based epidemiological model critically depends upon the realistic estimates of human mobility, which is often simulated using suite of  agent-based simulations, network science approaches, and data science methods \cite{eubank2004modelling,salathe2010high,zhang2017spread}. For example, Eubank et al. (2004) demonstrated the applicability of highly resolved agent based simulation tools that combine census and land use data with parameterized models to simulate the progression of infectious disease in realistic urban social networks \cite{eubank2004modelling}. Balcan et al. (2009) analyzed mobility data from various countries around the world and integrated in a worldwide structured meta-population mechanistic epidemic model to understand the role of infection due to multi-scale dynamic processes \cite{balcan2009multiscale}.
Ajelli et al.(2010) noted a good agreement between highly detailed agent-based modeling approaches and spatially structured mechanistic meta-population models. However, researchers note that while mobility networks used in meta-population models provide an accurate description of spreading phenomenon, detailed estimates of the impact at finer scales is hampered by the the low level of detail contained in such modeling schemes. On other hand, while agent-based modeling approaches are highly detailed, gathering high confidence detailed datasets, specifically in heterogeneous regions across the world, is a challenge \cite{ajelli2010comparing}. In the context of vector borne epidemics such as Zika virus (ZIKV) epidemic, researchers have integrated the high spatial and temporal resolution real-world demographic, mobility, socioeconomic and climatic condition datasets to estimate the profiles of ZIKV infections \cite{zhang2017spread}. In other studies, researchers have combined the state-space models such as population based Suspected (S), Exposed (E), Infected (I), and Recovery (R) or SEIR models in combination with Google Trend datasets to understand the evolution of Flu trends in the United States \cite{dukic2012tracking}. SARS-CoV-2 epidemic spread across the Globe in early 2020, mathematical, statistical and machine learning based models gained prominence to find out how to slow and or stop the spread \cite{gilbert2020preparedness,chinazzi2020effect,kraemer2020effect,vespignani2020modelling} . Moreover,  researchers referred  to the ongoing  efforts to model the spread of SARS-CoV2 as "war time"research where scientists have to deal with limited data, multiple assumptions and changing landscapes \cite{vespignani2020modelling}.  We note that despite the structural, mathematical and procedural differences which exist in the wide spectrum of epidemic models, realistic representation of human mobility patterns, and social interactions play a key role in governing model performance to understand the disease progression.  

To model human mobility,  travel demand models and  activity based models are typically used. While travel demand models focus on estimating aggregate road usage in long run using aggregated zonal statistics often obtained from census and household surveys. More recent methods try to learn about human behavior in urban areas by using data collected from location-aware technologies including telecommunication activity datasets and Global Position System data archives \cite{jiang2016timegeo}.We note that while massive and passive cellphone data can effectively generate complete urban mobility patterns \cite{hasan2013spatiotemporal}, these datasets are typically non-open source given privacy and security concerns. In absence of high resolution activity data, simplified models of human mobility such as  gravity models are often used to understand the spreading of viruses and the evolution of epidemics \cite{ajelli2010comparing,mari2012modelling,li2011validation}. 

The novel Coronavirus (2019-nCoV) first identified in Wuhan, China has spread to 213 countries as of August 2020 with more than 24 Million confirmed cases reported worldwide \cite{dong2020interactive}. While much still needs to be learnt about the virus, its clinical characteristics, extent of inter-human transmission , and the spectrum of clinical disease, it is established that transmission of SARS-CoV-2 occurred to great extent through superspreading events \cite{perlman_another_2020}. While accurate forecasting of spread as well as number of deaths and recoveries require ample historical data, contact networks, and identification of zeroth case in different regions in addition to the clinical parameters \cite{petropoulos2020forecasting}. While multiple modeling groups across the globes have proactively focused on modeling the spread at global and regional scales \cite{clark2020global,vespignani2020modelling,chinazzi2020effect}, studies at urban and city scales are limited \cite{minetto2020measuring,tuite2020estimation}. Given the diversity in socioeconomic factors, spread characteristics, demographics, healthcare facilities, management of 2019-nCoV needs to involve local negotiations. Moreover, while stringent lockdowns and travel restrictions were imposed by many nations, compliance with such preventive measures varies considerably from one to another region \cite{wright2020poverty,briscese2020compliance,painter2020political}. As a consequence, regions and cities with similar population densities and comparable socioeconomic indicators exhibited large disparities in the 
trajectory of the report SARS-CoV-2 cases.

In the present study, our objectives are two-folds. First, we attempt to understand the response of travel patterns in the city to partial lockdowns or creation of containment zones -- zones which are declared as no travel zones due to widespread detected cases in a particular area. Using network science based node prioritization approaches adapted from \cite{bhatia2015network} and travel demand models from \cite{ganin2017resilience}, we model the emergence of congestion zones within the city as a result of enforcement of containtment measures. We note that while creation of the containment zones are typically done to contain the spread of  SARS-CoV-2, emergence of the congestion zones as a consequence of traffic rerouting could facilitate the disease spread. 

Second, we attempt to understand the relationship between drivers of intra-city mobility and SARS-CoV-2 spread trajectory in different parts of the city.  We use generalized stochastic SEIRS infectious disease dynamic model that allow
us  to model the effect of social contact network structures, heterogeneities,  and interventions, such as social distancing, testing, contact tracing, and isolation on the spread trajectory of the disease.  We hypothesize that in addition to the population density, ease of transit within smaller regions, which in turn is facilitated by the shorter transit/trip times, can result in accelerated spread of the SARS-CoV-2. We simulate the mobility patterns in different regions of the city of Ahmedabad which have disparate road network distribution, land use patterns but comparable population sizes. We select the period of nationwide lockdown as the temporal window for analysis. Since  non-essential inter-district and interstate travel was prohibited during the period of lockdown, we simulate the travel patterns confined within the regions to mimic the population mobility for essential services.


\section*{Region of Study and Data}

We use Ahmedabad Municipal Corporation (AMC) as the study region (Figure \ref{fig:Figure 1}). Ahmedabad city has a population of 5.6 million (Census 2011), making it the fifth-highest populous city in India. The city's geographical location is 23.0225\textsuperscript{0} N, 72.5714\textsuperscript{0} E.  In the context of the SARS-CoV-2 outbreak, Ahmedabad city witnessed a sudden increase in the patient count since the appearance of first case on March 19,2020  and thereafter has been classified as one of the hotspot cities\cite{srivastava2020geographical}. 

The flowchart of the methodology is shown in (Figure \ref{fig:Figure 2}). The Road data in the form of a shapefile is extracted from an open street map (OSM) data sets to generate a transportation network. India's latest census data is from the year 2011, which may not provide an accurate result due to the duration gap. To address this, we considered the modeled population data of the year 2020 provided by the open source portal (www.worldpop.org) with a spatial resolution of 30 arc-second (approximately 100m). To understand the city's traffic pattern, we use the Traffic Demand Analysis report and database which is publicly available at  \cite{gidc}. The SARS-CoV-2 patient time series data with location is collected from COVID India Tracker- an open-source repository that compiles the patient data from various government sources in near-real time \cite{covid19indiaorg2020tracker}.

\section*{Methods and Materials}
\subsection*{Geo-spatial boundary and Population assignment}
We converted the road network into an undirected network graph, which translates the road's geometries into nodes and links. The attributes extracted from the data are the nodes' location, length of links, categories of links, and the links' end coordinates. The travel time of the road section is a crucial parameter to understand the behavior of traffic. We integrated travel time considering the category and length of a road section.

To define the transportation network boundary in  AMC,  we used the AMC boundary geo-spatial data and extracted the modeled population data for the same (Figure \ref{fig:Figure 3}(a)). To include the incoming and outgoing traffic from the city's outskirts, we generated a buffer radius of 1km. We considered the incoming traffic on the boundary junction of the city network (Figure \ref{fig:Figure 3}(c)). Then the whole boundary of the city is split into a unique region of the Voronoi cell (Figure \ref{fig:Figure 3}(d)). Then we estimated the population intersecting with each unique cell (Equation \ref{eq:Equation-1}). Each cell population data is transferred to their respective road junctions.
\begin{equation}\label{eq:Equation-1}
 P_{i} = \Sigma\: P_{t} \times \frac{Area\:(C_{i}\times C_{t})}{Area\:C_{t}} 
\end{equation} 
   
   where:
   \begin{itemize}
			    \item $P_{i}$ = Population in each Voronoi polygon 
				\item $P_{t}$ = Modeled population 
				\item $C_{i}$ = Cell of Voronoi polygon 
				\item $C_{t}$ = Cell of population 
				
			\end{itemize}

We also incorporated the land use (Figure \ref{fig:Figure 3}(b)) into the analysis to create real traffic congestion scenarios for post-lock down situations in the city. The assumption is made that around 1/4\textsuperscript{th} of the population will commute from residential to commercial for daily work, making the industrial and commercial area more crowded compared to the residential area. The re-allocation of population assignment indicates the highly congested pockets in the city.

\subsection*{Traffic demand forecasting model }
The transportation model enables travel demand forecast while taking the population and household travel survey data into account. We generated a commuter pattern based on a  gravity model. The gravity model is a classical trip distribution model widely used for trip distribution in an urban context. Other approaches to model transportation networks, namely the queuing model\cite{Gross2013queuning}, growth factor model\cite{Easa1993growthmodel}, and activity based model using human mobility patterns \cite{gonzalez2008understanding}. These alternative approaches are not considered based on (a) methods are data extensive and  (b) some of the techniques required the observed data, which is very difficult to gather during the lockdown traffic patterns are completely different when compared to original traffic mobility.  We calculated the trip for road section based on the number of trips attracted to or leaving from specific zones. These trips are distributed by the linking of origin and destination, thus forming an Origin-Destination matrix. The allocation of these trips is done by considering the shortest possible route.  We used the Djikstra algorithm \cite{Dijkstra1959connexion} to find the shortest commuter distance. (Equation \ref{eq:Equation-2})

\begin{equation}\label{eq:Equation-2}
T_{ij} = P_{i} \times  \frac{\exp{(-aC_{ij})}P_j}{\sum_{(i,j) \in (V_S \times V_S)} P_j\exp{(-aC_{ij})}}
\end{equation}

			Where:
			\begin{itemize}
			    \item $T_{ij}$ = the number of trips produced in the zone i and attracted to zone j 
				\item $V_S$ = the set of road intersections in the region $S$
				\item $P_i$ = the population of the $i^{th}$ road intersection
				\item $C_{ij}$ = the minimum time required to go from $i^{th}$ road intersection to $j^{th}$ road intersection.
				\item a = adjustable parameter to be fixed after model calibration
			\end{itemize}

\subsection*{Network Analysis}
To understand the network structure's intricate pattern and relationship, we carried out the network analysis using a network's centrality measures . Centrality provides a vital and widely used measure to analyze the network as it helps determine the most critical nodes in the network \cite{bhatia2015network}. Betweenness centrality is an important criterion for analyzing centrality, especially in road networks. Road networks include commuters' flow along its edges and assuming that people tend to optimize their commute path by taking the shortest route. We calculated betweenness centrality for each junction  in the network and ranked according to this centrality value (Equation \ref{eq:Equation-3}). The ranks demonstrate the junction's vulnerability in the network and demarcate the tendency for high connection with other junctions and can be considered as a measure for the potential for the junction to develop into congestion points. We assigned the trip count calculated from the gravity model to each road section (links) and ranked them based on the possible transition. This provides the section with the highest vulnerability in terms of increased traffic flow.
\begin{equation}\label{eq:Equation-3}
     Cb (v) =\sum_{u \in v}  \sum_{w \in v/{u}}\frac {\sigma_{uw} (v)}{\sigma_{uw}}  
\end{equation}   
    	Where:
			\begin{itemize}
			    \item $\sigma_{uw}$ = the number of shortest paths between u and w 
				\item $\sigma_{uw} (v)$ = the number of shortest paths between u and w that passes v
			\end{itemize}

\subsection*{Scaling of disease spread}
We use the derived transportation network to estimate the spreading of disease in selected regions with comparable population within the cities. The spread of disease is dependent on population density. The scaling of disease spread to a city level helps to understand the behavior of spread in the urban environment. Population density is used to estimate disease scaling, the relation between them is not always linear. Moreover, this estimate holds true when the extrapolation is done over a small region as the exposure is dependent on the social interaction of the people. Hence, these estimations based on a linear relationship between population density and disease scaling work effectively when there is  uniformity in the community. 

On a city scale, though, the extent of such predictions is extremely limited due to the vast diversity in living conditions and connectivity contrast in different regions. Here we have assumed that people's interaction in a region is more if the connectivity inside is competently developed. Based on the above-mentioned criteria, we calculated the scaling factor to remove a few limitations on the population density based extrapolation. The scaling factor for a region $S$ with respect to another region $S_0$, $\kappa(S, S_0)$ is defined in (Equation \ref{eq:Equation-4}).

\begin{equation}\label{eq:Equation-4}
\kappa(S, S_0) = \frac{N_S \log{N_S}}{N_{S_0} \log{N_{S_0}}} \times \frac{A({S_0})}{A(S)} \times \frac{F({S_0})^2}{F(S)^2}
\end{equation}

\begin{itemize}
			\item $N_S$ = Population of region $S$
			\item $A(S)$ = Area of region $S$
			\item $F(S)^2$ = Interaction Coefficient
		\end{itemize}

Here the F (S) is the independent interaction coefficient of the region and it is calculated using (Equation \ref{eq:Equation-5}).
\begin{equation}\label{eq:Equation-5}
	F(S) = \sum_{(i,j) \in (V_S \times V_S)} \frac{\exp{(-C_{ij})}P_i P_j}{N_S^2}
\end{equation}
			
			Where:
			\begin{itemize}
				\item $V_S$ = the set of road intersections in the region $S$
				\item $P_i$ = the population of the $i^{th}$ road intersection
				\item $C_{ij}$ = the minimum time required to go from $i^{th}$ road intersection to $j^{th}$ road intersection in hours.
			\end{itemize}
			
The route that gives the minimum time, $C_{ij}$ is calculated based on the assumed travel speed for disparate categories of roads. The complete calculation can be done using the transport network mentioned in the previous subsection. The value of $F(S)$ is always less than 1 and the square of this value gives the interaction coefficient for the region $S$. The exact details for this construction are given in the supplementary section.

The scaling factor we have defined in the Equation \ref{eq:Equation-4} is a relative quantity between region $S$ and $S_o$. To provide a single value that can be used to directly quantify multiple regions and to allow direct comparison between them, we use a base region for all calculations. The base region is the identity region $I$, which is not physically present but an imaginary construct such that $\frac{N_I log N_I}{A(I) F(I)^2} = 1$. Using this identity region as the base, for any two regions $S_1$ and $S_2$ (Equation \ref{eq:Equation-6})
\begin{equation}\label{eq:Equation-6}
			\kappa(S_1, S_2) = \frac{\kappa(S_1, I)}{\kappa(S_2, I)}
\end{equation}			
			For notation purpose, we define $\kappa(S) = \kappa(S, I) = \frac{N_S log N_S}{A(S) F(S)^2}$
			
It is to be noted that the interaction factor is inversely proportional to the scaling factor. This is because the interaction factor does not consider the area of the region, which is separately considered in the calculation. The construction of the scaling factor is explained in a supplementary section.

\subsection*{SEIR PLUS model}

SEIR PLUS  epidemic spread model is applied to scaled-up data to predict the trajectory of spread. The SEIR model is a standard compartment based model. We used the dynamic form of the model in this study, which is used on stochastic dynamic networks \cite{SEIRpluse_model}.
The equations governing the state of the nodes is explained in (Equation \ref{eq:Equation-7}).

\begin{equation}\label{eq:Equation-7} 
		    Pr(X_i = S \rightarrow E) = \left[ p\frac{\beta I}{N} + (1-p)\frac{\beta \sigma_{j\in C_G(i)} \delta_{X_j=I} }{|C_G(i)|} \right] \delta_{X_i=S}
\end{equation}

		    $$Pr(X_i = E \rightarrow I) = \sigma \delta_{X_i=E}$$

            $$Pr(X_i = I \rightarrow R) = \gamma \delta_{X_i=I} $$

		    $$ Pr(X_i = I \rightarrow F) = \mu_I \delta_{X_i=I} $$

		    $$ Pr(X_i = R \rightarrow S) = \eta \delta_{X_i=R} $$

For testing purposes and quarantining, the above equations can be modified by adding some more compartments to account for quarantining and testing. The details are discussed in a supplementary section.

\subsubsection*{The clinical parameters}
Parameters such as the rate of transmission, rate of progression, recovery rate, mortality rate depend on the cause of the epidemic. We obtained these parameters from the clinical data \cite{Godio2020SEIR}\cite{Bagal2020SEIRIndia}. At the same time, there is also a dependency on the recovery rate when a positive person is detected early on or late.

We assumed that these parameters are independent of other factors related to social interaction and lockdown effects. A range was determined using available clinical data for such parameters, and fixed values from this range were used throughout.

\subsubsection*{Setting up the Scenarios}

The interventions brought by the Government brings changes in the interaction of people compared to daily life. At the same time, a person detected positive will reduce the interaction with the rest of the population. To model the different scenarios, we use Erd\H{o}s-R\'eyni graphs with different probabilities for edge creation. These probabilities are seen in the form $\alpha/N_S$ for a region  $S$.

\subsubsection*{Tuning of parameters}
The tunable parameters are the testing rates at different periods. Also, the SEIR model requires the number of initially infected people.

\subsection*{Calibration of epidemic spread model}

We now detail the steps followed during the calibration cycle:

\begin{itemize}
    \item Pick a single sub-region, say $S^1_5$, and run the SEIR plus model with some arbitrary plausible values for the initial infected parameter, graph $\alpha$ values, and testing rates.
    \item Using the fitting between the observed data for infection and the resuts of the SEIR model to fine-tune the values to be used for testing rates and initial infected rates.
    \item At the same time, change the $\alpha$ parameters but keep them same across all regions.
    \item Check the output for all $10000$ population subregions. Do further tuning of $\alpha$ values for the different garphs and also fine tune the testing parameters and initial infected for each region. It must be kept in mind that the initail infected are set only for $5000$ population subregions and scaled values are used for $10000$ population subregions.
    \item Find a optimal way such that all plots fit with the observed values for all $5000$ and $10000$ population subregions
    
\end{itemize}

\section*{Results}
 Our transportation model has two indicators: Betweenness Centrality rank (BC rank) of a road junction (Figure \ref{fig:Figure 4}(a)) and probable trip count of each road section(Figure \ref{fig:Figure 4}(b)). These indicators help to identify densely connected pockets in the city. Through plotting these results, We have demarcated the possible traffic congestion pockets in the city. These results show the city pockets vulnerable to spread in post-lockdown scenarios.  Our analysis shows that Ahmedabad's central region has high BC rank, demonstrating the priority intersection that can be vulnerable to disruption due to heavy traffic. The road segments with the high trip count are located in the city's central and eastern parts, where population density is high. Also, it was seen that the first cases of COVID -19 were observed in this part of the city, making the transportation model useful while considering drive through testing. 

We simulated targeted disruption in the road network to analyze the system's response, which resembles the often seen situations in SARS-CoV-2 spread. Due to many cases in a particular region, the region might get quarantined or declared a containment zone. The containment zones make up the proper condition in network science that is termed as targeted disruption. The city network is witnessing the interruption in the traffic flow with the declaration of these containment zones as no travel zones. To analyze the disruption, we removed one of the city's containment zones. The possible rerouting shows that this triggers the new potential vulnerable region to traffic congestion and spread through our re-calibrated indicators (Figure \ref{fig:Figure 5}).

The prediction of disease spread is difficult in the diverse living conditions and connectivity. In that case, we hypothesize that the scaling factor derived from the transportation network model provides a better alternative to population density. It considers the population density and the interaction between the community by road intersections. The travel time plays a crucial role in determining the scaling factor. To understand the importance of travel time, we have considered two regions with the same populations (Figure \ref{fig:Figure 6}(a)). The travel time changes with the amount of intersections in the region.That is lesser number of intersections in a a region would translate to longer transit times, on an average, in an urban setting. This eventually leads to a decrease in the interaction coefficient and an increase in the scaling factor. Using this concept, the population density is calculated from population data (Figure \ref{fig:Figure 6}(b)) and the scaling factor through interaction coefficient (Figure\ref{fig:Figure 6}(c)) is calculated. Subsequently, we check the hypothesis by determining the Kendall-Tau Coefficient generated between the cumulative ward wise infected cases and the scaling factor and population density, respectively (Figure\ref{fig:Figure 6}(d)). This analysis validates our hypothesis and shows that the scaling factor provides a better correlation compared to the population density (Figure \ref{fig:Figure 7}(a)) and  \ref{fig:Figure 7}(b)). 

Once the hypothesis is tested, the calibrated clinical parameters are used to validate the prediction of infected population for the regions. The epidemic SEIR plus model implemented on sub-regions with disparate ranges of population such as $5000$, $10000$ and $20000$ is used to validate the model parameter values (Figure \ref{fig:Figure 8}(a), (b) and (c)). This approach allows us to account for the variation that comes up with different issues of testing and variation in actual testings when only average testing rates are known. To check the appropriateness of the model the time observations of duration considered in the testing are the ones, which are not considered during the calibration by plotting calibration and testing plot simultaneously.  For the region 1 with 20000 population, the SEIR plus model predicted infected population shows the good agreement with the observed data. Region-1 shows the R\textsuperscript{2}, Relative Root Mean Square Error (RRMSE) and Relative Mean Absolute Error (RMAE) value as 0.9602, 0.688 and 0.558 respectively (Figure \ref{fig:Figure 8}(d)).  The same approach we have implemented in region 2, region 3 and region 4 respectively (Figure \ref{fig:Figure 9}). We evaluate the prediction performance for the above mentioned region and it is shown in (Table \ref{tb:Table 1}). The value of R\textsuperscript{2} is smaller in the region 4 as the data for SARS-CoV-2 infected cases in the region was less compared to other regions.

The SARS-CoV-2 spread has led governments to think of different interventions to reduce the spread by implementing various lockdown policies. These policies can change the trajectory of disease spread. To account this into action, we have also simulated the disparate policies in our model by selecting social distancing and relaxation combinations.

 The first policy considered is that if the government implements the lockdown relaxation by last week of May 2020, Unlock with the strict social distancing till mid-June 2020 and then staged back to a normal state in 50 days. The result shows that we can expect a sudden spike of cases in the initial days after lockdown relaxation. However, it gradually decreases in the upcoming months, but it shows the second wave of cases forming from September 2020 (Figure \ref{fig:Figure 10}(a)). 
 
The second policy we consider is the lockdown reduced by May 2020 end, Unlock with the strict social distancing followed for 15 days and then staged back to normal state by mid-July 2020 in a staggered manner.  The model result demonstrates increased infected cases in the initial days after lockdown relaxation and shows the fluctuation in the infected cases in the September and October 2020 (Figure \ref{fig:Figure 10}(b)).

Lastly, the third policy simulated is that the lockdown reduced by May end, Unlock with the strict social distancing followed for one month and then staged back to a normal state. The result demarcates that there is a rise in cases in the initial few days, but after that, from the month of July-2020, it can drop down very low (Figure \ref{fig:Figure 10}(c)).

\section*{Discussion}
The prediction of disease spread at the city scale is often complex due to diverse regional factors and data limitations. The road networks are the prime sources of intracity movements. This movement pattern can lead us towards the possible spreading of disease as most of the infectious diseases spread through social interactions. During these scenarios, prediction of disease spread through network-based epidemic spread models can be very helpful. This study has presented the unique approach to model the disease spread through transportation networks. The result of this analysis allows several meaningful inferences, which can make a high impact on the prediction of disease spread. First, this approach can be implemented in any congested city to determine the interaction between the population and lead to model disease spread. Most of the data considered during the modeling are open source or readily available. Second, framework also provides the flexibility of understanding disparate scenarios such as containment zone restriction. The third inference gained from the analysis is that the different lockdown policies are highly influenced by social interaction and can be analyzed through the network-based epidemic spread modeling. Policymakers can choose the best possible way to contain the spread. Although the interaction will vary from city to city based on local conditions, we anticipate that the overall patterns will be similar for comparable population densities and road networks. We further note that the quality, quantity and frequency of epidemiological datasets play an important role in establishing the correlations that we have observed in this study. While the proposed framework can be generalized to other cities, future efforts in this direction can greatly benefit from real-time mobility data obtained from cellphone activity or GPS data, and high-resolution clinical and epidemiological data with relatively longer duration of record.

%
%
%



\begin{backmatter}

\section*{Competing interests}
  The authors declare that they have no competing interests.

\section*{Author's contributions}
    RP,UB, RD  designed the experiments; RP, UB, RD, HP, VS performed the analysis; RP, RD, UB, and DC wrote the manuscript. 

\section*{Acknowledgements}
  This work is supported by Startup Research Grant awarded to UB by Indian Institute of Technology Gandhinagar. We are thankful to Prasanna V Balasybramanium, Assistant Professor, Indian Institute of Technology, Gandhinagar, for his helpful comments and suggestions. \ldots

\bibliographystyle{bmc-mathphys} 
\bibliography{bmc_article}      




\section*{Figures}
\begin{figure}[h!]
    \centering
    \includegraphics[scale=0.6]{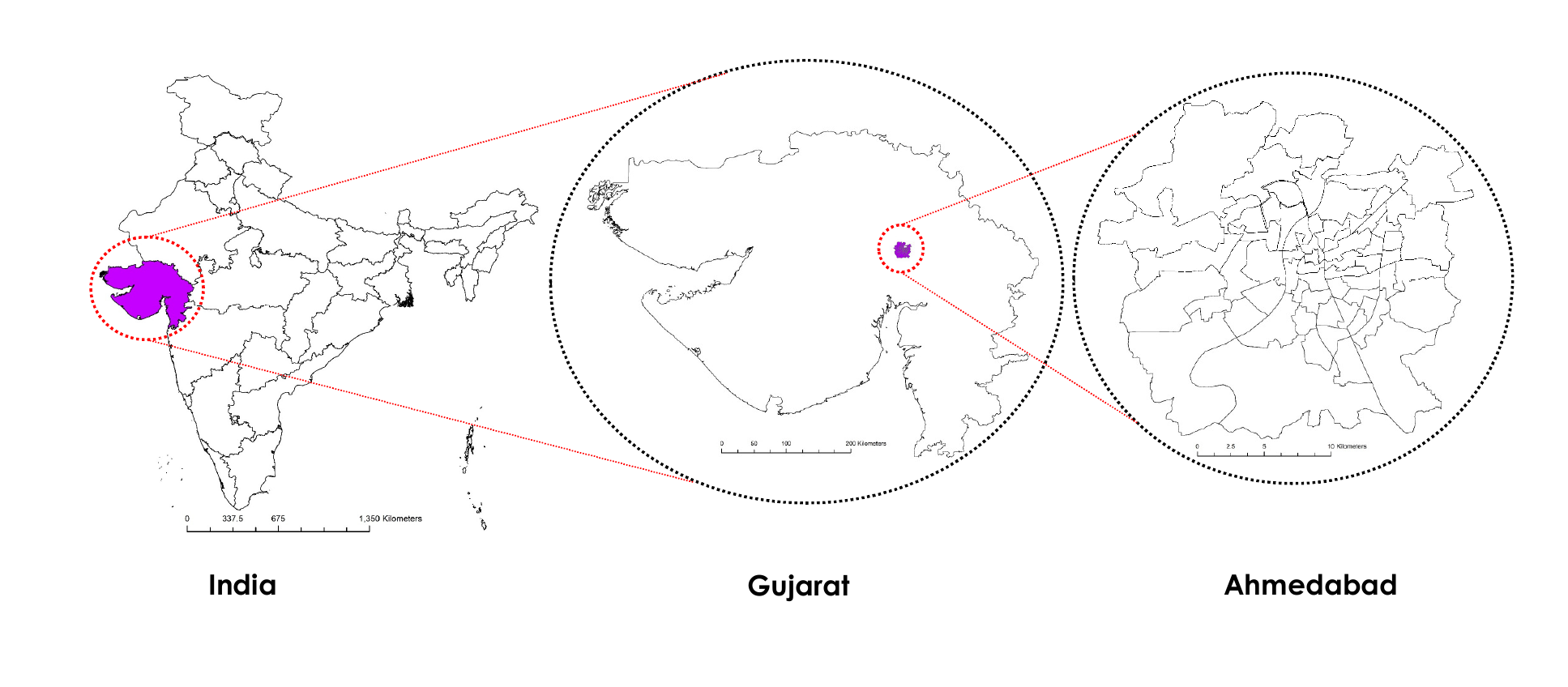}
    \caption{ Study area of Ahmedabad Municipal Corporation (AMC), located in the central region of Gujarat in western India. Ahmedabad is selected as area of study given its distinction as financial hub of Gujarat, disparate socioeconomic distribution and high population density }
    \label{fig:Figure 1}
\end{figure}

\begin{figure}[h!]
    \centering
    \includegraphics[scale=0.6]{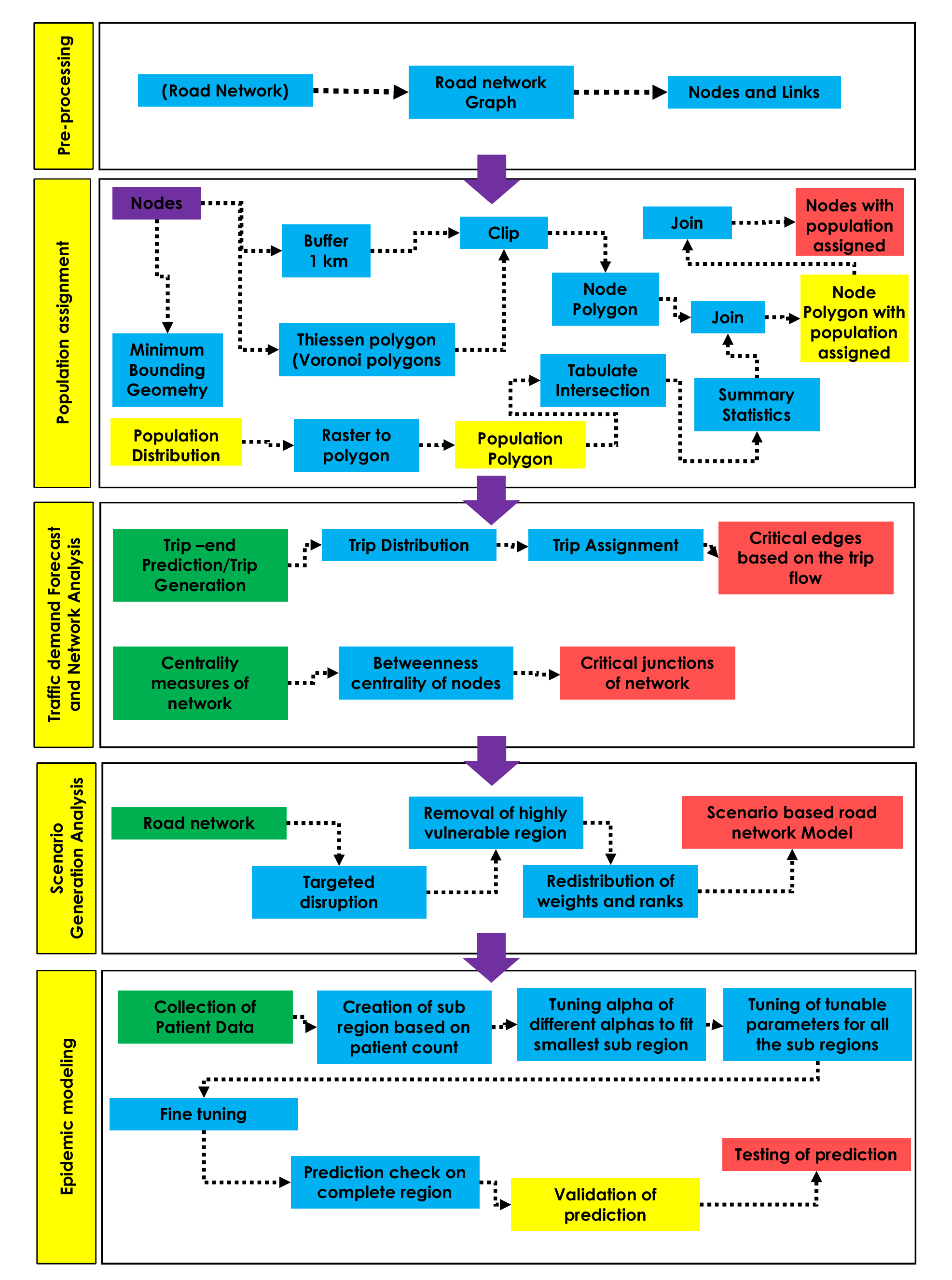}
    \caption{Overview of the modeling framework. The proposed methodology consist of (a) Pre-processing, (b) Population assignment, (c) Traffic demand forecast and network analysis, (d) Scenario generation analysis (Targeted disruption) and (e) Epidemic spread modeling}
    \label{fig:Figure 2}
\end{figure}

\begin{figure}[h!]
    \centering
    \includegraphics[scale=0.6]{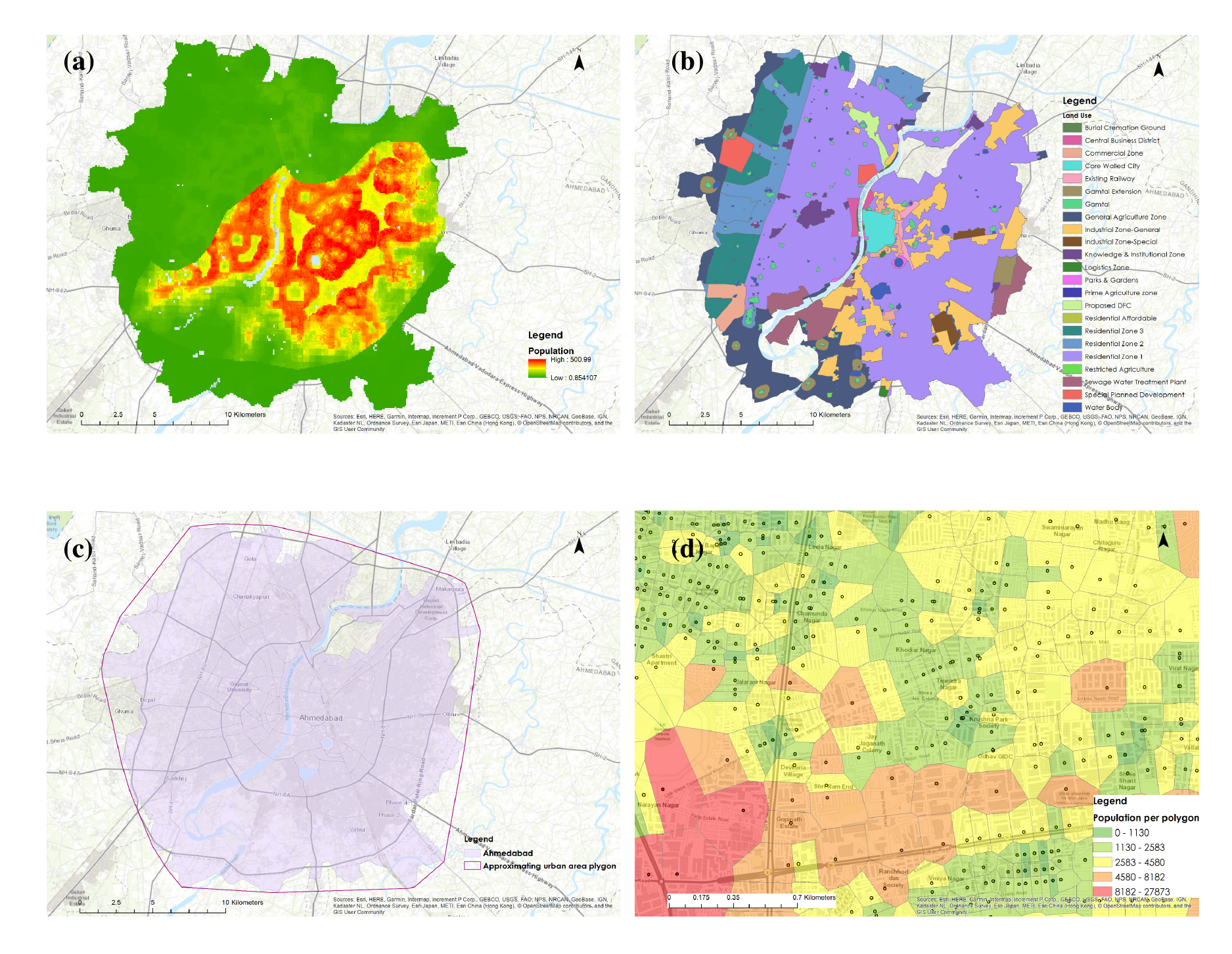}
    \caption{(a) Modeled population data extracted at 30 arc-second , (b) Land use distribution describing diverse arrangement of land use in the city ,(c) Bound polygon to consider incoming and outgoing traffic and (d) Assignment of number of population served by junction in a unique Voronoi polygons.(red colour is resemblance of high population, where as green colour is for low population}
    \label{fig:Figure 3}
\end{figure}

\begin{figure}[h!]
    \centering
    \includegraphics[scale=0.6]{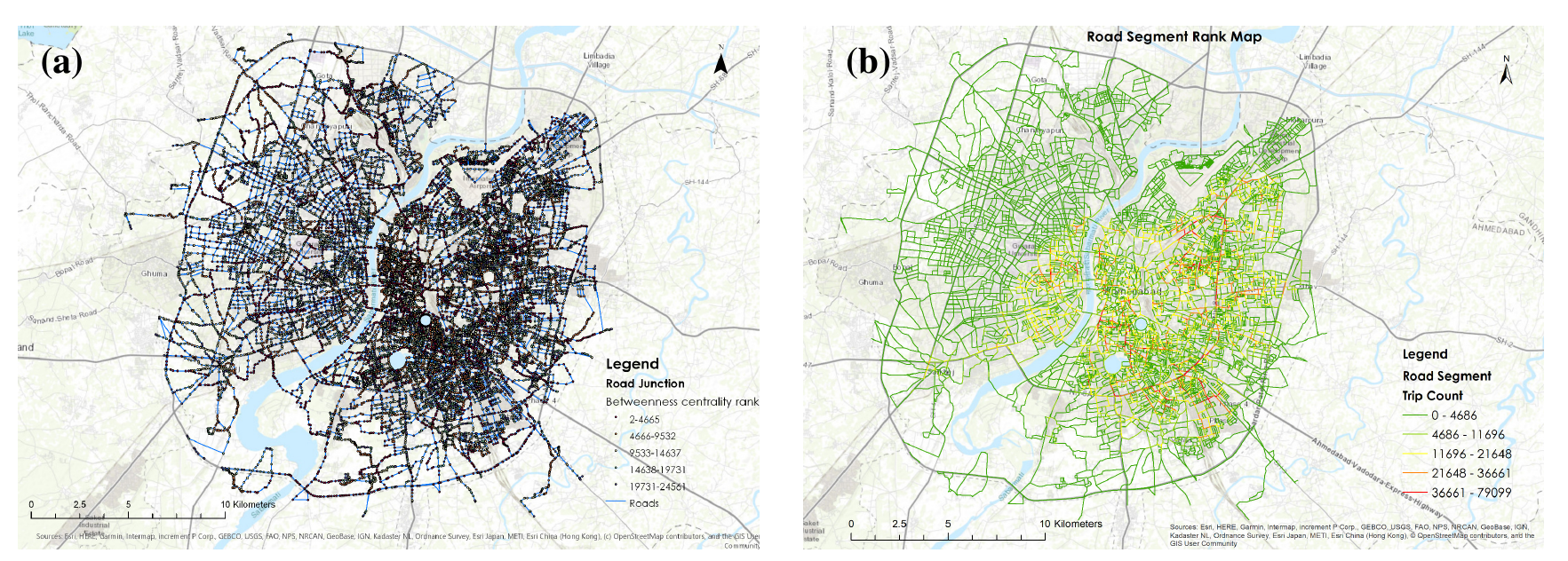}
    \caption{Maps of Ahmedabad city depicting  (a) Road junction distribution according to BC Rank (Red colour nodes depicting high BC rank) and (b) Road network link distribution according to trip count (Red colour road sections depicting high trip count)}
    \label{fig:Figure 4}
\end{figure}

\begin{figure}[h!]
    \centering
    \includegraphics[scale=0.6]{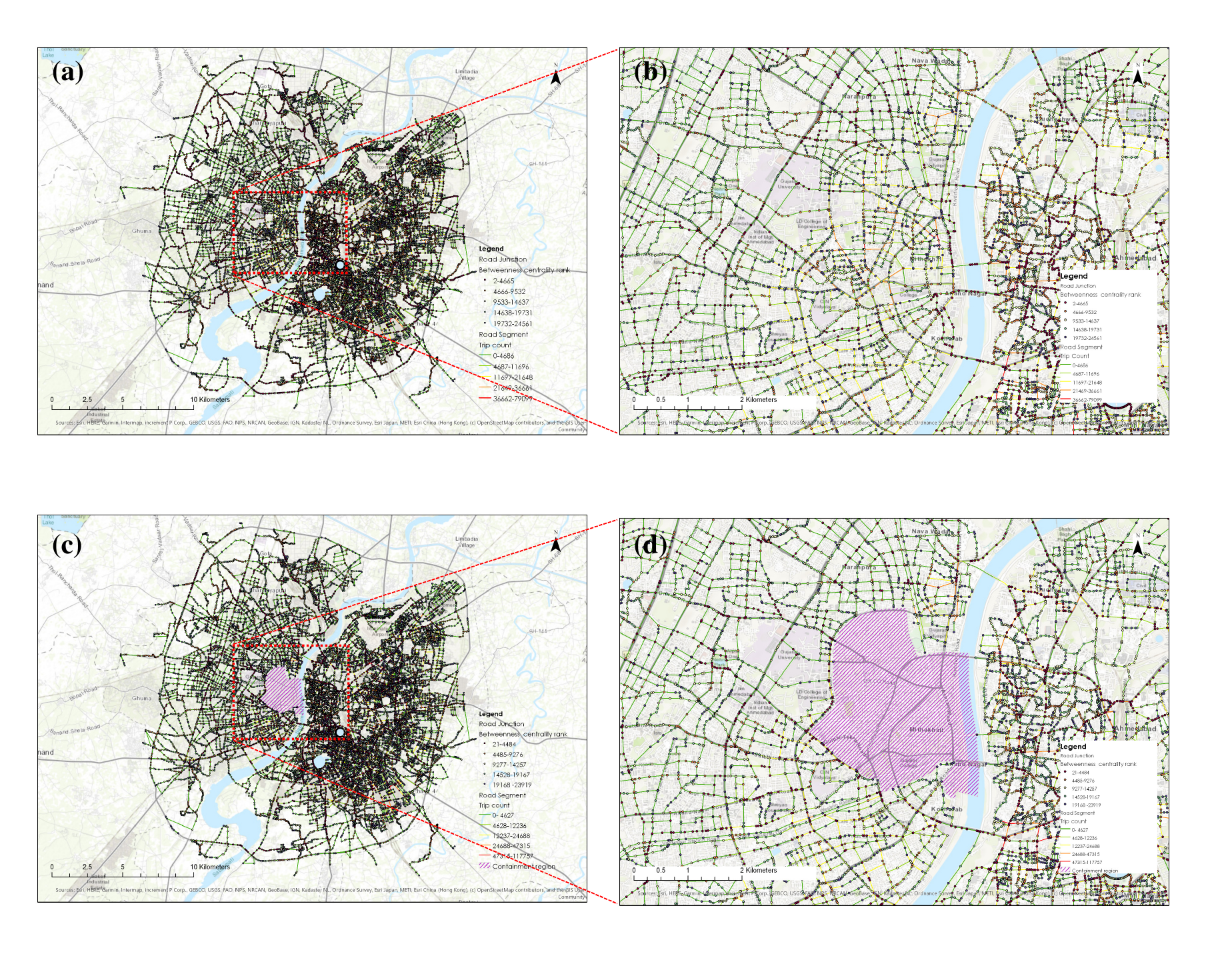}
    \caption{(a) The road network without disruption (normal condition), (b) Zoomed in map of road network without disruption, (c) Targeted disruption of road network to generate the lockdown scenario by removing containment zone and (SARS-CoV-2 spread situation) (d) Zoomed in map of targeted disruption}
    \label{fig:Figure 5}
\end{figure}

\begin{figure}[h!]
    \centering
    \includegraphics[scale=0.6]{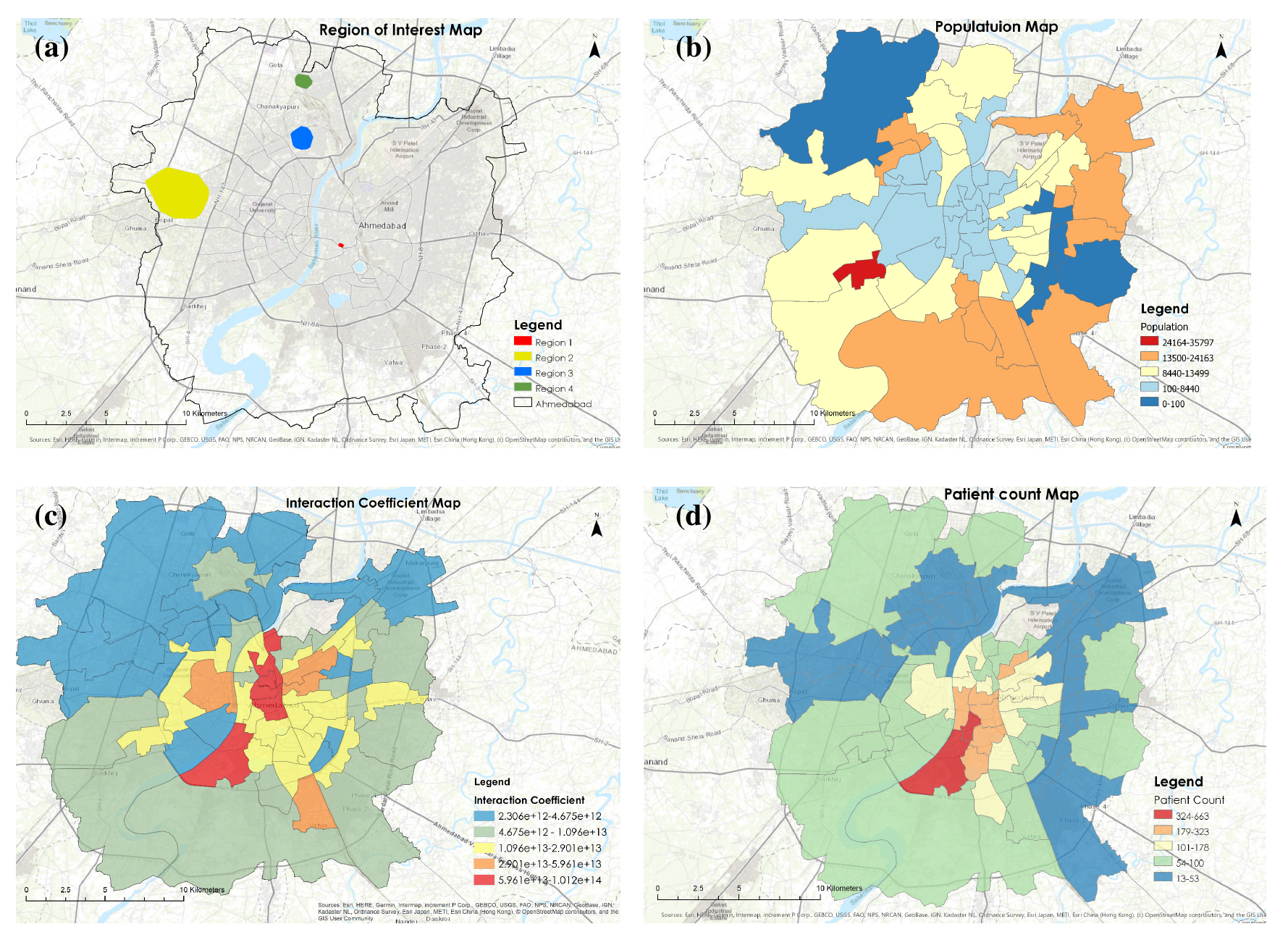}
    \caption{(a) Region of interest in the study area for epidemic spread model with same population,(b) Population distribution in the city, (c) Interaction Coefficient distribution in the city and (d) SARS-CoV-2 Patient Distribution}
    \label{fig:Figure 6}
\end{figure}

\begin{figure}[h!]
    \centering
    \includegraphics[scale=0.6]{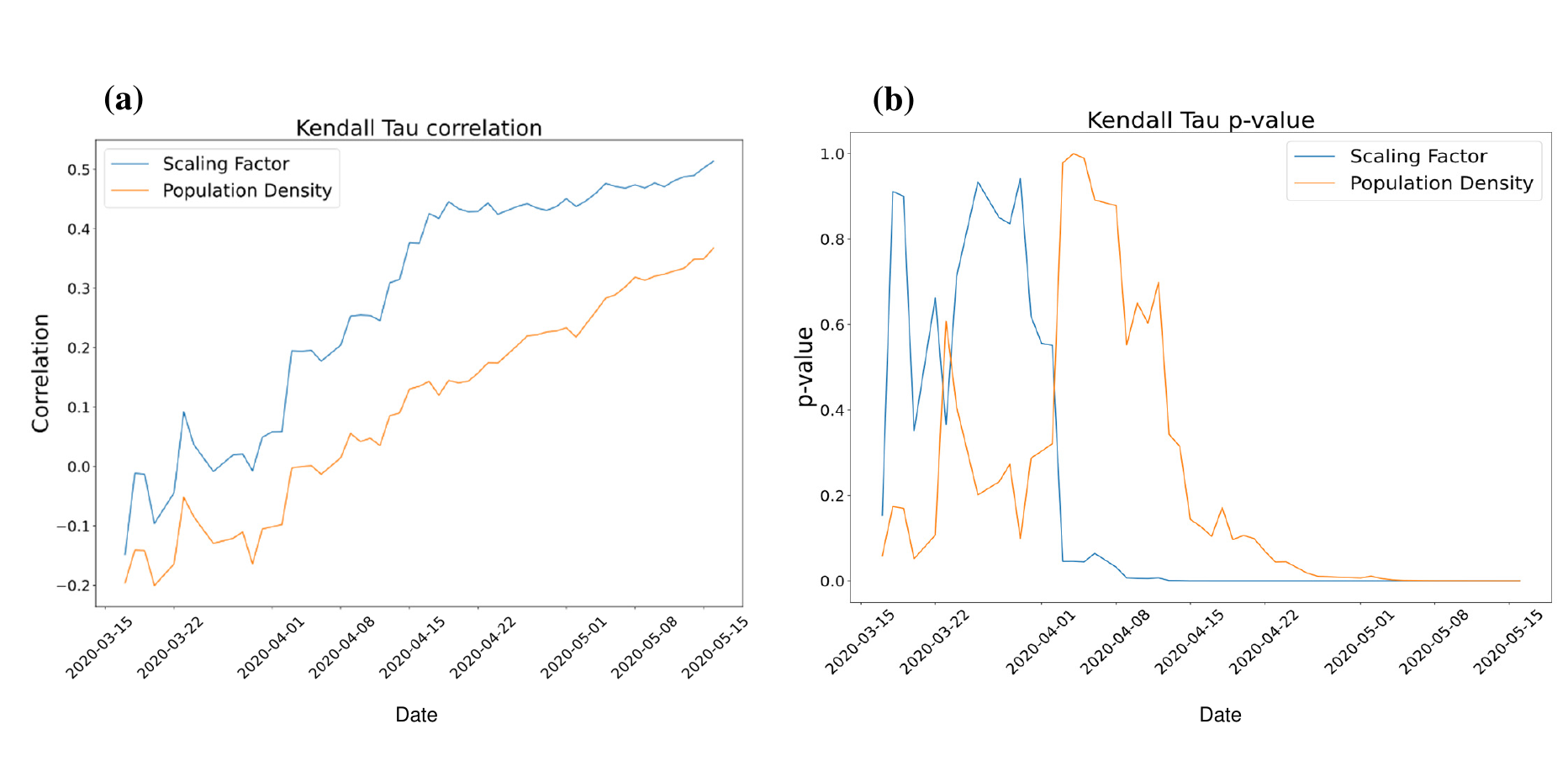}
    \caption{(a) Kendall Tau Correlation and (b) Kendall Tau p- value}
    \label{fig:Figure 7}
\end{figure}

\begin{figure}[h!]
    \centering
    \includegraphics[scale=0.6]{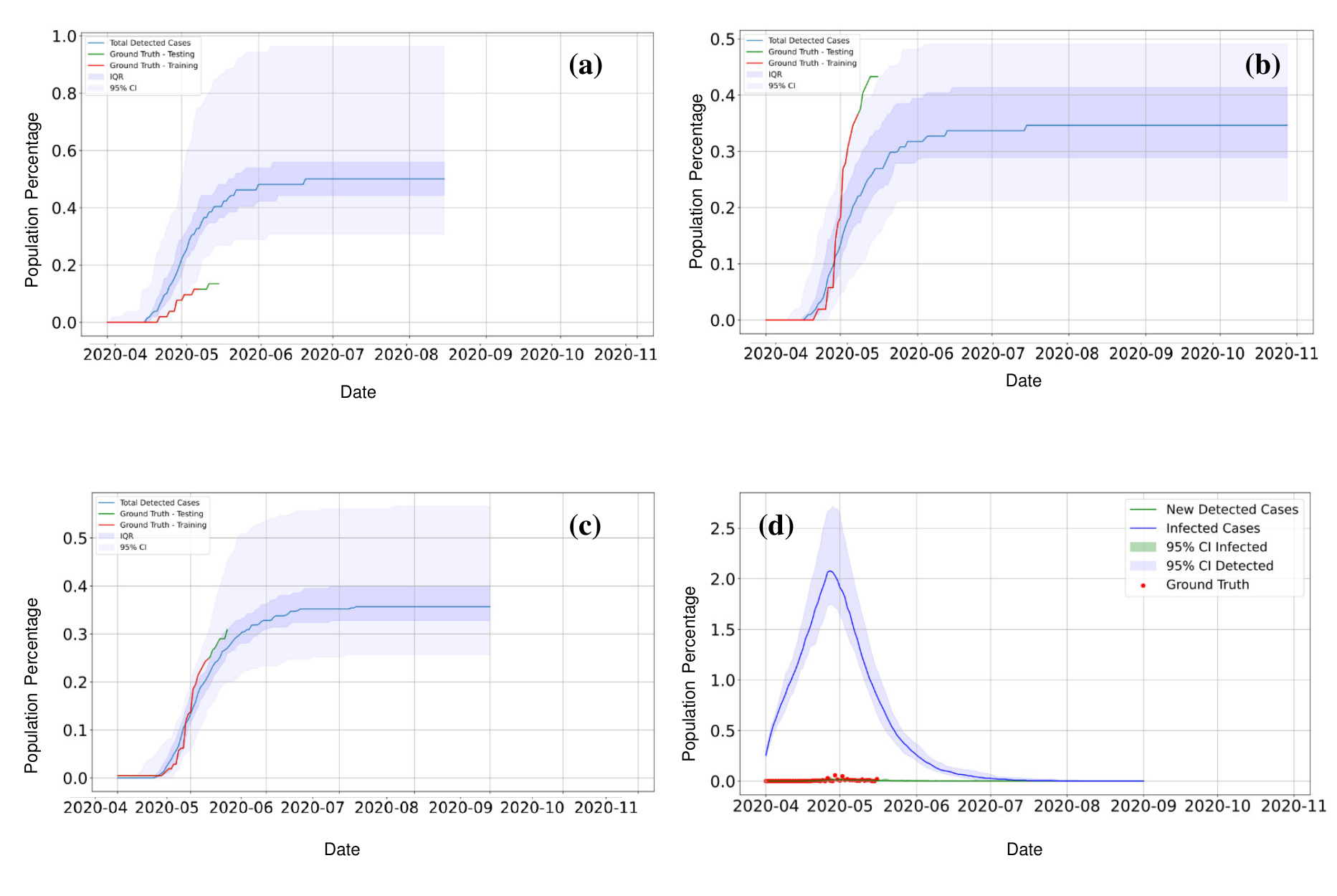}
    \caption{Region 1 time series data set: (a) Training Data set (1-5000 Population cumulative cases, (b) Training Data set (1-10000 Population cumulative cases), (c) Validation Data set (1-20000 Population cumulative cases) and (d) Prediction Data set (1-20000 Population cumulative cases) }
    \label{fig:Figure 8}
\end{figure}

\begin{figure}[h!]
    \centering
    \includegraphics[scale=0.6]{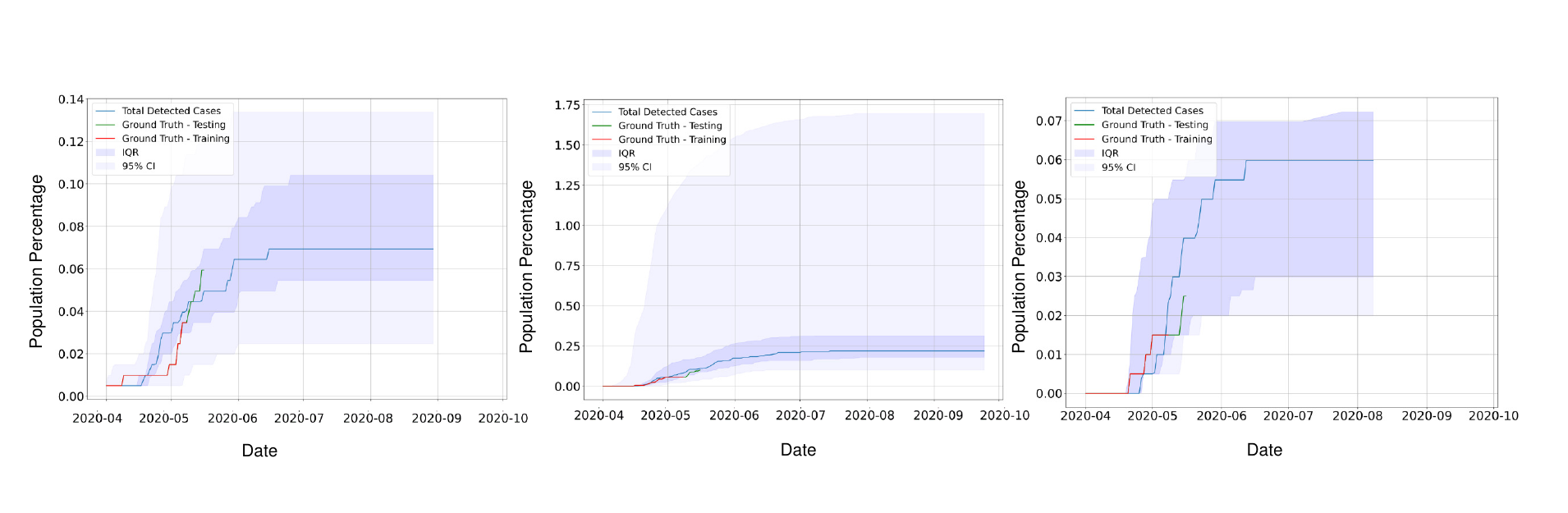}
    \caption{(a)Testing data sets: Region 2 Data set (1-20000 Population cumulative cases), (b) Region 3 Data set (1-20000 Population cumulative cases) and (c) Region 4 Data set (1-20000 Population cumulative cases)}
    \label{fig:Figure 9}
\end{figure}
\clearpage
\begin{figure}[h!]
    \centering
    \includegraphics[scale=0.6]{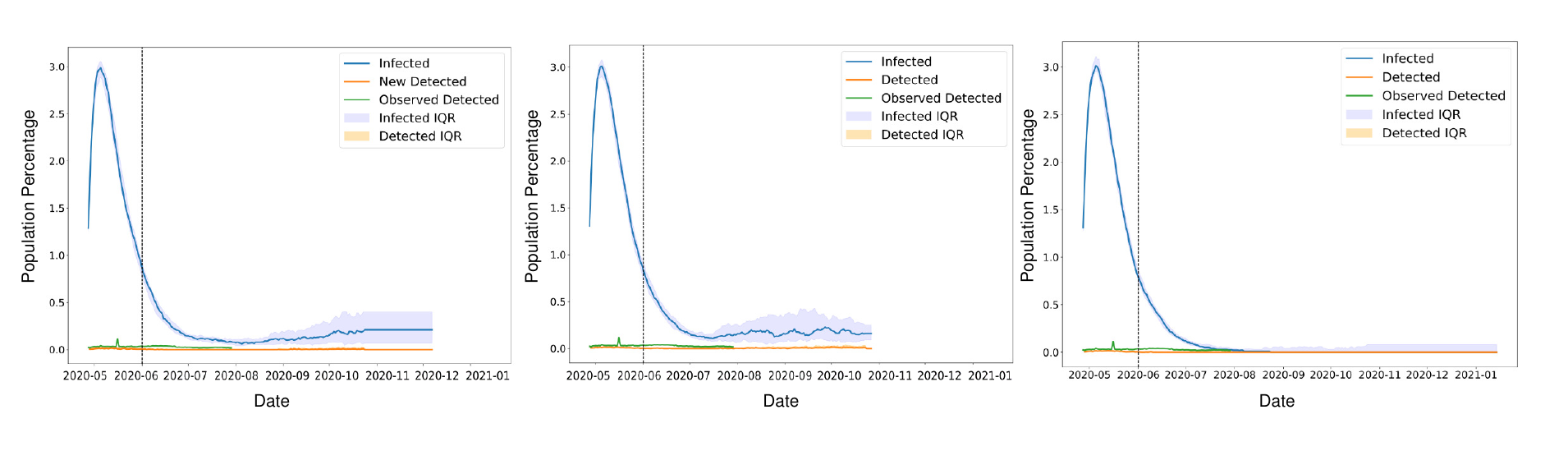}
    \caption{Time series data of patient count based on different lock down policies (a) Lock down relaxation by last week of May, Unlock with the strict social distancing till mid June and then staged back to normal state in 50 days , (b) Lock down reduced by May end, Unlock with the strict social distancing followed for 15 days and then staged back to normal state by mid July in staggered manner, (c) Lock down reduced by May end, Unlock with the strict social distancing followed for 1 month and then staged back to normal state}
    \label{fig:Figure 10}
\end{figure}


\section*{Tables}
\begin{table}[h!]
\caption{Error matrix of prediction depiction the model performance on region of interests}
        \begin{tabular}{|c|c|c|c|}
        \hline
        \textbf Regions (20000 population) & \textbf R\textsuperscript{2} & \textbf RRMSE & \textbf RMAE\\ \hline
        1 &  0.96 & 0.68 & 0.55\\
        2 & 0.68 & 0.73 & 0.47\\
        3 & 0.87 & 0.41 & 0.30\\
        4 & 0.15 & 0.71 & 0.64\\ \hline
        \end{tabular}
        \label{tb:Table 1}
\end{table}


\section*{Supplementary Materials}
  \subsection*{Interaction coefficient}
  We constructed an interaction coefficient based on the gravity model. The interaction between the two intersections is determined based on the two parameters: The first one is the fraction of the population in the individual intersections and the second is the time required to travel from one intersection to another. $C_{ij}$ gives the information of the time needed by a person while taking the best path possible for travel.
  
  $C_{ij}$ is considered zero when $i$ and $j$ intersection are the same. For which the average time required for the interaction between two persons is believed to be very small. This holds when $C_{ij}$ is measured in terms of hours or days. This assumption breaks when the measurement is in smaller units as the intra-intersection interaction cannot be approximated to $0$.
  
 Considering the constraint mentioned above, the calculation of $C_{ij}$ is done in terms of hours. The selection of larger units such as day or month is not considered in the analysis because for the intersection in proximity distance, the travel time will become negligible, leading to a single intersection. In the city environment,  a person can reach any part of the city and nearby places in the time interval of hours. The unit of time for $C_{ij}$, in a sense, it defines the resolution at which the system is being observed.  If we consider the single interaction in the region $ S $, it is trivial to see that $F(S) = 1$. In the case of two intersections, the value of $F(S)$ will decrease as $C_{ij} \neq 0$ for all terms. The more the skewed the distribution of population in the region, the smaller is the independent interaction coefficient.  
  
  $F(S)$ only considers the independent interaction between two intersections. But a commuting person may have to go from one intersection to another always through some other intersection. For example, consider that there are $3$ intersections in a region $S$.( Equation \ref{eq:Equation 8})
 \begin{equation}\label{eq:Equation 8}
     F(S) = \alpha_1^2+\alpha_2^2+\alpha_3^2+2(c_{12}\alpha_1\alpha_2 + c_{23}\alpha_2\alpha_3 + c_{13}\alpha_1\alpha_3)
\end{equation} 
    where $\alpha_i$ is the fraction of population for $i^{th}$ intersection and $c_{ij} = \exp{(-C_{ij})}$.
    
 considering (Equation \ref{eq:Equation 8}) the interaction coefficient is 
    \begin{equation}
    \begin{split}
    	F(S)^2 & = \alpha_1^4 + \alpha_2^4 + \alpha_3^4 + 4(c_{12}^2\alpha_1^2\alpha_2^2 + c_{23}^2\alpha_2^2\alpha_3^2 + c_{13}^2\alpha_1^2\alpha_3^2) \\
    	       & \quad + 4(c_{12}\alpha_1^3\alpha_2 + c_{23}\alpha_1^2\alpha_2\alpha_3 + c_{13}\alpha_1^3\alpha_3) \\
    	       & \quad + 4(c_{12}\alpha_1\alpha_2^3 + c_{23}\alpha_2^3\alpha_3 + c_{13}\alpha_1\alpha_2^2\alpha_3) \\
    	       & \quad + 4(c_{12}\alpha_1\alpha_2\alpha_3^2 + c_{23}\alpha_2\alpha_3^3 + c_{13}\alpha_1\alpha_3^3) \\
    	       & \quad + 8(c_{12}c_{23}\alpha_1\alpha_2^2\alpha_3 + c_{12}c_{13}\alpha_1^2\alpha_2\alpha_3 + c_{13}c_{23}\alpha_1\alpha_2\alpha_3^2) \\
    	       & = \alpha_1^4 + \alpha_2^4 + \alpha_3^4 + 4(c_{12}^2\alpha_1^2\alpha_2^2 + c_{23}^2\alpha_2^2\alpha_3^2 + c_{13}^2\alpha_1^2\alpha_3^2) \\
    	       & \quad +  4(\alpha_1^2+\alpha_2^2+\alpha_3^2)(c_{12}\alpha_1\alpha_2 + c_{23}\alpha_2\alpha_3 + c_{13}\alpha_1\alpha_3) \\
    	       & \quad + 8c_{13}\alpha_1\alpha_2\alpha_3(\alpha_2 + c_{12}\alpha_1 + c_{23}\alpha_3)	      \\ 
    \end{split}
    \label{eq: Equation 9}
    \end{equation}

The last term merges and accounts for the path that is passing through the 2\textsuperscript{nd} intersection. If we extend it to $n$ intersections, there will be many cross-terms that will be extra. Also, there will be terms like $c_{12}c_{34}$, where the question comes that how did a person jump from 2\textsuperscript{nd} intersection to 3\textsuperscript{rd} intersection. A person may not always go by the shortest path but take small detours in the way. Suppose the shortest path from 1\textsuperscript{st} intersection to 4\textsuperscript{th} was direct, but the person went in 1-2-3-4 path. Then a single cross term cannot represent this information and the cross-terms with jumps account for this to some extent.

Calculating the population's interaction considering multiple paths that can be taken between two is an exponentially tricky question to solve. Taking the square of the independent interaction coefficient provides a simple approximation for the same.

\subsection*{Scaling Factor}
    The inverse proportionality of the scaling factor to the interaction coefficient comes because it captures the skewness of interaction in a fixed population of fixed area.If we consider two regions with the same population and area, one with a single intersection and one with two intersections. For simplicity, in the second region, assume that the individual intersections are associated with half the area. In that case, if there is some travel time between them, then we cannot say that the population is using half of the area in the region. The population are using less area then that and this depends solely on the travel time given that both intersections have the same population. This demarcates that more people are packed in the smaller region and the overall interaction will fall, while the the intra-intersection interaction will be more and the scaling of spread will increase.
    
    Similarly, suppose only one intersection contains all the population and there is a significant travel time between the two intersections. In that case, this implies that most people are packed inside a small region, leading to a large decrease in interaction factor and a shoot in the scaling factor. This odd behavior can also be seen from the fact that the interaction factor normalizes on the basis that every node in a region has internal interaction proportional to the square of its population fraction.  To correctly scale with the area and population of the region, $\kappa(S)$ is multiplied with $\frac{N_S}{A(S)}$ which is the population density of the region.
    
    According to the Erd\H{o}s-R\'eyni model for random graphs, the hard threshold for the connectedness of a graph $G(n, p)$ is $\frac{\log{n}}{n}$. The graphs mentioned in (Setting up the Scenarios section) will require the $\log{N_S}$ factor for scaling the probability correctly for the graphs as interaction coefficient and population density together quantify the net increase in connectedness of the random graph for that region.

\subsection*{SEIRS Plus with testing and quarantine}

We add two more compartments, $D_E$ and $D_I$, representing the detected exposed and detected infected. To model the quarantine, a quarantine graph $Q$ is required, where the assumption is that detected people are in quarantine, and to model the testing, the parameters $\theta_E$, $\theta_I$, $\psi_E$ and $\psi_I$ are added, where $\theta_E$ and $\theta_I$ are the testing rates for exposed and infected people respectively and $\psi_E$ and $\psi_I$ are the rate of the positive test result for exposed and infected individuals respectively. The $\theta$ are tunable parameters, while $\psi$ are clinical parameters \cite{SEIRpluse_model}.

\begin{equation}\label{eq:Equation 10}
		    Pr(X_i = S \rightarrow E) = \left[ p\frac{\beta I + q\beta_D D_I}{N} + (1-p) \left( \frac{\beta \left[ \sigma_{j\in C_G(i)} \delta_{X_j=I} \right] + \beta_D \left[ \sigma_{k\in C_Q(i)} \delta_{X_k=D_I} \right] }{|C_G(i)|} \right) \right] \delta_{X_i=S}
		\end{equation}
		
		   $Pr(X_i = E \rightarrow I) = \sigma \delta_{X_i=E}$

		   $Pr(X_i = I \rightarrow R) = \gamma \delta_{X_i=I}$

		  $Pr(X_i = I \rightarrow F) = \mu_I \delta_{X_i=I}$

		  $Pr(X_i = R \rightarrow S) = \eta \delta_{X_i=R}$

		    $Pr(X_i = E \rightarrow D_E) = \left( \theta_E + \Phi_E \left[ \sigma_{j\in C_G(i)} \delta_{X_j=D_E} + \delta_{X_j=D_I} \right] \right) \psi_E \delta_{X_i=E}$

		   $Pr(X_i = I \rightarrow D_I) = \left( \theta_I + \Phi_I \left[ \sigma_{j\in C_G(i)} \delta_{X_j=D_E} + \delta_{X_j=D_I} \right] \right) \psi_I\delta_{X_i=I}$

		    $Pr(X_i = D_E \rightarrow D_I) = \sigma_D \delta_{X_i=E}$
		
		    $Pr(X_i = D_I \rightarrow R) = \gamma_D \delta_{X_i=I}$

		    $Pr(X_i = D_I \rightarrow F) = \mu_D \delta_{X_i=I}$

		    $Pr(X_i = any \rightarrow S) = \nu \delta_{X_i\neq F}$
		
	Where:
	    \begin{itemize}
	        \item $\nu$: Rate of baseline birth
	        \item $\mu_D$: Rate of infection-related mortality for detected cases
	        \item $\sigma_D$: Rate of progression for detected cases
	        \item $\gamma_D$: Rate of recovery for detected cases
	        \item $\beta_D$: Rate of transmission for detected cases
	        \item $\Phi_E$: Rate of contact tracing testing for exposed individuals
	        \item $\Phi_I$: Rate of contact tracing testing for infected individuals
	    \end{itemize}

\end{backmatter}
\end{document}